\documentclass[pdflatex,sn-mathphys-num]{sn-jnl}

\usepackage{graphicx}%
\usepackage{multirow}%
\usepackage{amsmath,amssymb,amsfonts}%
\usepackage{amsthm}%
\usepackage{mathrsfs}%
\usepackage[title]{appendix}%
\usepackage{xcolor}%
\usepackage{textcomp}%
\usepackage{manyfoot}%
\usepackage{booktabs}%
\usepackage{algorithm}%
\usepackage{algorithmicx}%
\usepackage{algpseudocode}%
\usepackage{listings}%
\usepackage{xspace}
\usepackage{subcaption}

\newcommand{\s}{\ensuremath{\sqrt{s}}\xspace}
\newcommand{\ups}{\ensuremath{\Upsilon}(1S)\xspace}
\newcommand{\py}{PYTHIA~8\xspace}
\newcommand{\jpsi}{J/\ensuremath{\psi}\xspace}
\newcommand{\qqbar}{\ensuremath{Q\bar{Q}}\xspace}

\newcommand{\gevc}{\ensuremath{\rm{GeV/}c}\xspace}
\newcommand{\pt}{\ensuremath{p_{\mathrm{T}}}\xspace}
\newcommand{\oldpy}{PYTHIA~8.307\xspace}
\newcommand{\newpy}{PYTHIA~8.312\xspace}
\newcommand{\z}{\ensuremath{z}\xspace}

\theoremstyle{thmstyleone}%
\theoremstyle{thmstyletwo}%

\theoremstyle{thmstylethree}%

\begin{document}

\title[Article Title]{J/$\psi$ and $\Upsilon$(1S) production in jets at LHC energies}


\author*[1]{\fnm{Lizardo} \sur{Valencia Palomo}}\email{lizardo.valencia.palomo@cern.ch}

\affil*[1]{\orgdiv{Departamento de Investigaci\'on en F\'isica}, \orgname{Universidad de Sonora}, \orgaddress{\street{Blvd. Luis Encinas y Rosales S/N, Col. Centro}, \city{Hermosillo}, \postcode{83000}, \state{Sonora}, \country{Mexico}}}


\abstract{Quarkonia production in hadronic collisions is far from being understood as non of the existing models can correctly describe the wealth of available data. In particular, LHCb and CMS experiments have reported that PYTHIA 8 can not reproduce the prompt \jpsi production in jets in proton-proton collisions at two different center of mass energies: the event generator predicts an important amount of the prompt \jpsi to be produced isolated, opposite to the experimental data. This document demonstrates that such effect remains true even if the QCD color reconnection (CR) model is used. Besides that, it is shown that using the new quarkonia parton shower included in \newpy it is possible to correctly describe the experimental results. This agreement between data and simulation is improved when using the QCD color reconnection approach, opening the possibility to distinguish between the two CR implementations. Finally, a prediction performed for \ups indicates that a higher jet \pt selection should be used by the LHC experiments in order to distinguish between PYTHIA 8 results generated with and without the quarkonia parton shower.}


\maketitle

\section{Introduction}\label{sec1}

Quarkonia are hadrons made up of a heavy quark, charm or beauty, and its corresponding anti-quark. If the bound state is a pair of charm (beauty) and anti-charm (anti-beauty) quarks then it is called chamonium (bottomonium). Charmonia states can be further divided into prompt and non-prompt. The former refers to those directly produced in the collision, while non-prompt are those originating from the decay of beauty hadrons. 

According to Quantum Chromodynamics (QCD), quarkonia production in hadronic collisions can be split into three well defined stages \cite{FactorizationTheo}. The first one falls into the domain of non-perturbative QCD, as it addresses the probability of finding a parton within the colliding hadrons with a given fraction of momentum. The second stage corresponds to the hard scattering and is related to the partonic cross section for the heavy quark ($Q$) and heavy anti-quark ($\overline{Q}$) pair production. This is the only stage that can be computed using perturbative QCD. Finally, there is the evolution of the \qqbar pair into a physical quarkonium state that is also a non-perturbative QCD process. There are different models on quarkonia production and they mainly differ in the last step: the underlying mechanism that transforms a pre-resonance state of heavy quark and anti-quark into a hadron.

The Color Singlet Model (CSM) was the first to be proposed, right after the \jpsi was discovered \cite{CSM0,CSM1,CSM2}. This model is meant to describe the production process for $^3S_1$, singlet $S$, $P$ and $D$ states. Where the usual spectroscopic notation has been used: $^{2S+1}L_J$ with $S$ the spin, $L$ the orbital angular momentum and $J$ the total angular momentum. As the name suggests, the main assumption of the model is that the \qqbar pair emerging from the partonic scattering is directly produced as a color singlet state. 

Opposite to CSM, the Color Evaporation Model (CEM) makes no assumption on the color nor angular momentum quantum numbers of the pre-resonance state generated in the perturbative interaction \cite{CEM1,CEM2,CEM3}. For this reason, the \qqbar pair can be in a color singlet or color octet state. When the pre-resonance is produced in a color octet state it neutralizes the color by emission of a gluon: a sort of \textit{color evaporation}, giving birth to the name of the model. 

Due to their large masses, heavy quarks can be treated non-relavistically. For this reason the effective field theory called Non-Relativistic QCD (NRQCD) employs an expansion in terms of the velocity of the heavy quark or anti-quark in the quarkonium rest frame ($v$) to describe the hadronization of the \qqbar pair into a quarkonium state via Long Distance Matrix Elements (LDME) \cite{NRQCD1,NRQCD2}. So, quarkonia production cross sections depend on an infinite number of unknown matrix elements. In practice this infinite series can be truncated at a fixed order of $v$, in such a way that only a finite number of matrix elements are needed for the calculation. NRQCD makes no restriction on the color state of the pre-resonance so this can be produced as a color singlet or color octet, its spin can be singlet or triplet and it can also have angular momentum. As a consequence the corresponding truncation for each S-wave multiplet that includes color singlet and octate states is given by four matrix elements: $^3S_1^{(1)}$, $^1S_0^{(8)}$, $^3S_1^{(8)}$ and $^3P_J^{(8)}$. The extra superscripts (1) and (8) indicate if the color state is a singlet or octet, respectively. 

At the LHC both \jpsi and \ups production have extensively been studied in proton-proton (pp) collisions by the four main experiments at different center of mass energies \cite{CMSquarkonia,LHCbJpsi,LHCbUpsilon,ALICEquarkonium,ATLASjpsi}. But not only production cross sections have been reported, the wealth of data has favored precise studies of already known observables and even the emergence of new measurements that further constraint theoretical models \cite{CMSpolJpsi,LHCbPolUps,ALICEjpsiMult,CMSupsMult,ATLASupsMult}. Indeed, LHCb was the first to study the \jpsi production in jets at \s = 13 TeV in pp collisions \cite{LHCbJpsiJet}. Measurements as a function of $z$ (jet transverse momentum fraction carried by the quarkonium state) are compared to \py, indicating that for prompt J/$\psi$ the event generator clearly overestimates the data when $z \approx 1$. CMS has also performed similar measurements at \s = 5.02 TeV and once again the prompt component from the Monte Carlo overshoots the data when $z \approx 1$ \cite{CMSJpsiJet}. This means that the leading order NRQCD based prediction from PYTHIA 8 shows that prompt J/$\psi$ are produced with a small degree of surrounding jet activity, albeit the experimental results indicate a different trend

\section{Color reconnection and quarkonia production in PYTHIA 8}\label{sec2}

PYTHIA 8 is a general purpose Monte Carlo event generator used to describe high energy collisions between leptons, nucleons, pions, photons and nuclei \cite{Pythia8.3}. Some parts of the underlying physics are extracted from first principles, while some others rely on parameters tuned directly from experimental data. For pp collisions the event generator starts picking the initial partons from the Parton Distribution Function (PDF) of the incoming protons. Then the hard scattering of the two colliding partons generates outgoing particles whose kinematics are computed perturbatively. This hard scattering is not only accompanied by initial and final state radiation (parton shower) but also by Multiple Partonic Interactions (MPI): additional scatterings from partons present in the protons. In \py the confinement effect is represented by color singlets called strings or clusters, while their fragmentation into hadrons follows the Lund string model. But strings do not fragment independently, overlapping and interacting strings effects are described by a rope hadronization scheme. Finally, unstable particles decay, the products can re-scatter or recombine among them, and this process continues until only stable particles remain.

Partons produced along the collision are initially taged by color following the limit where the number of colors is very large, implying that different MPI processes are independent from each other. Color Reconnection (CR) allows string length minimization, so partons from unrelated MPI are color connected. In this work two of the different CR modes are applied: the MPI and QCD based. In the former model, the probability that two MPI get reconnected is governed by one free parameter called Reconnection Range (RR). According to this model low \pt partons are more likely to be color reconnected and if the value of RR increases, then more parton reconnections can be made. In the QCD based approach the string length minimization is further combined with SU(3) color algebra, providing a more realistic model \cite{QCD-basedCR}. The QCD color combinations create junction structures: topologycal compositions where three color lines meet at a single point, implying a richer colored topology than the MPI method. This CR mode adds a new mechanism for baryon and anti-baryon pairs creation. 

Concerning quarkonia production in hard scatterings, \py implements a NRQCD approach with only unpolarized processes. At the leading order, \qqbar production is due to gluon fusion ($g\,g \rightarrow Q\bar{Q}$) and light quark and anti-quark annihilation ($q\,\bar{q} \rightarrow Q\bar{Q}$). There is also the flavour excitation ($Q\,g \rightarrow Q\,g$) that occurs for the hard scatter and in the timelike QCD parton shower quarkonia can be produced by gluon splitting ($g \rightarrow Q\bar{Q}$) \cite{TimeLikePartonShower}. For $^3S_1$ states, like \jpsi and \ups, the even generator provides the corresponding color singlet and octet LDME for $^3S_1^{(1)}$, $^1S_0^{(8)}$, $^3S_1^{(8)}$ and $^3P_0^{(8)}$ states.

\begin{figure}
  \begin{subfigure}{0.49\textwidth}
    \includegraphics[width=\linewidth]{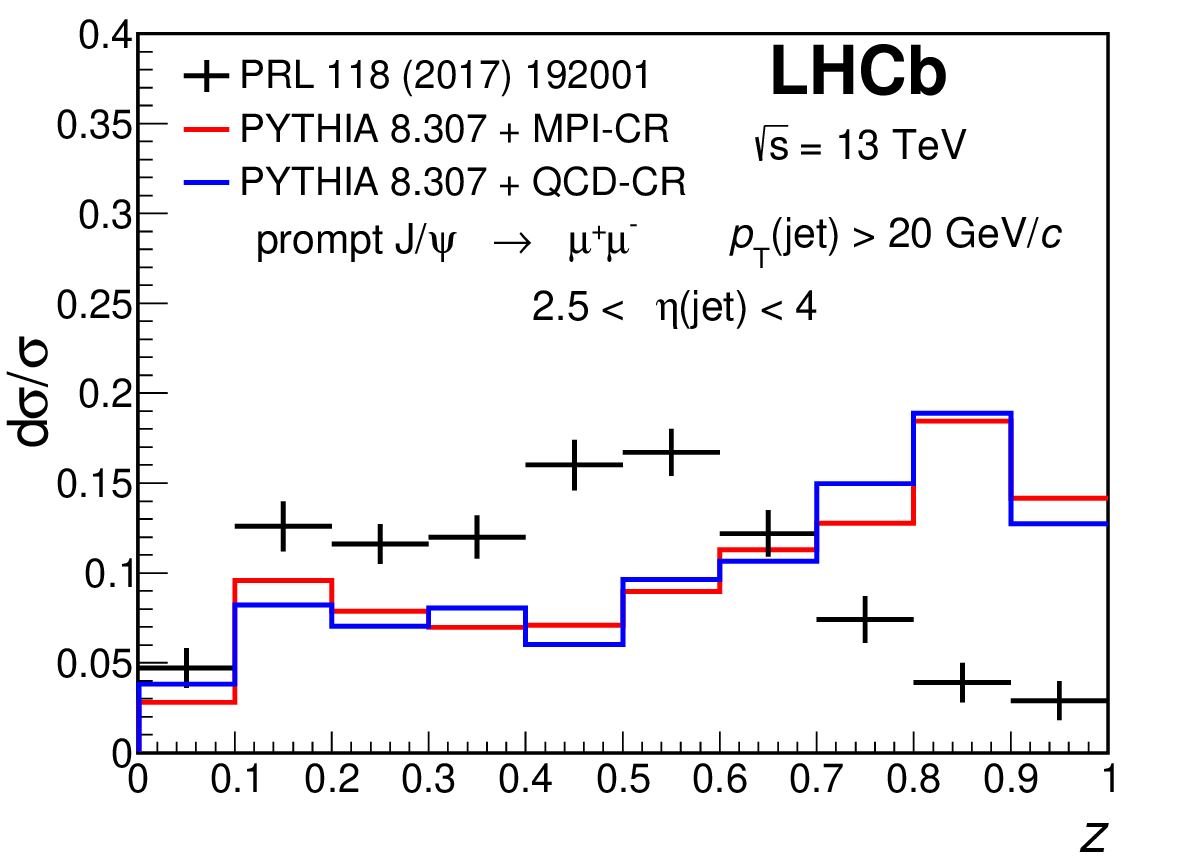}
    \caption{} \label{LHCbOriginal}
  \end{subfigure}
  \hspace*{-0.4cm}
  \begin{subfigure}{0.49\textwidth}
    \includegraphics[width=\linewidth]{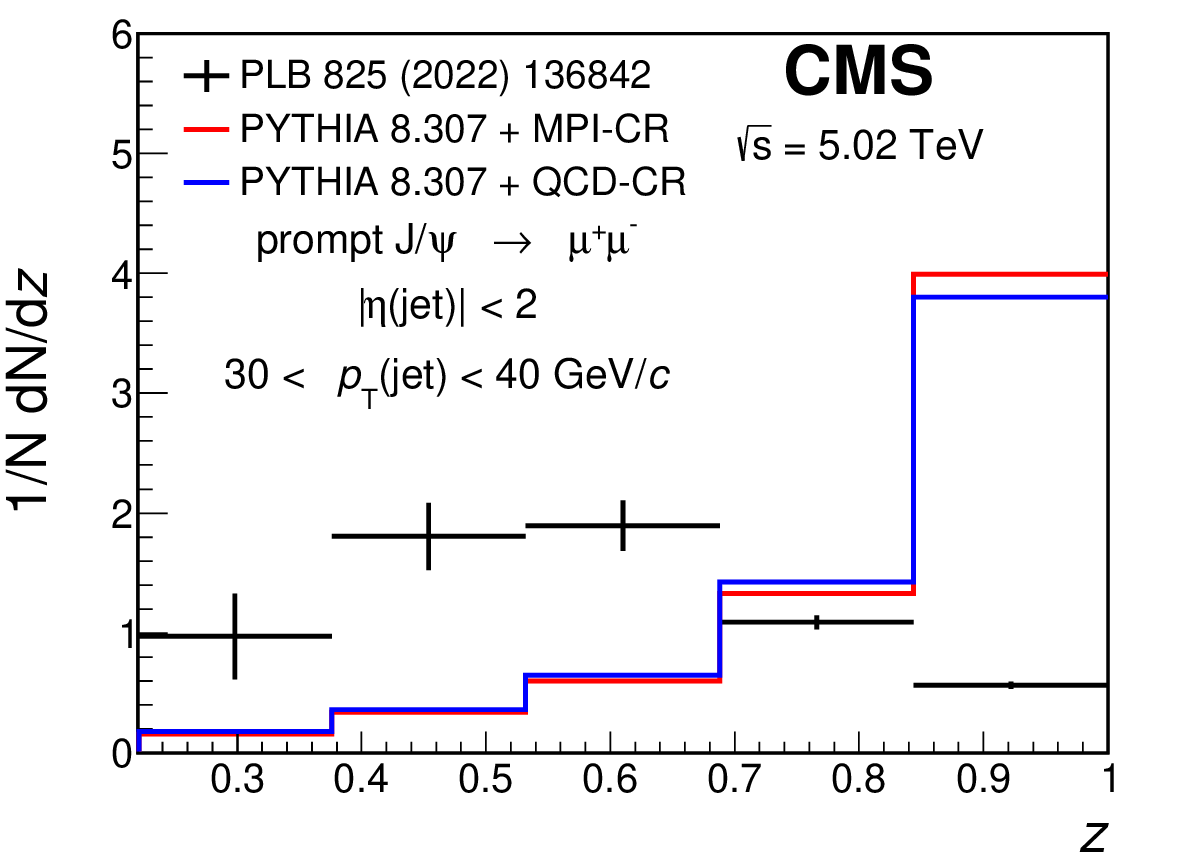}
    \caption{} \label{CMSOriginal}
  \end{subfigure}
\caption{Normalized \z distributions for prompt \jpsi in pp collisions at \s = 13 and  5.02 TeV as reported by LHCb (a) and CMS (b), respectively. For LHCb the error bars are systematic uncertainties only, while for CMS the statistical and systematic uncertainties have been added in quadrature. Solid lines are the predictions from PYTHIA 8.307, red line employs the MPI color reconnection approach while the blue one the QCD based scheme} 
\label{OriginalPlots}
\end{figure}

Figure \ref{LHCbOriginal} presents the fragmentation of jets containing a prompt \jpsi meson in pp collisions at \s = 13 TeV as a function of the jet transverse momentum fraction carried by the \jpsi (\z) as reported by LHCb \cite{LHCbJpsiJet}. This measurement is performed using \jpsi with \pt $>$ 0 and reconstructed using the dimuon decay channel. The anti-$k_{\mathrm{T}}$ algorithm from FastJet is used for the jet clustering requiring a radius $R=0.5$ and a \pt(jet) $>$ 20 \gevc at forward rapidity (2.5 $< \eta\mathrm{(jet)} <$ 4) \cite{AntiKtAlgo,FastJet}. Notice the \jpsi, not the decaying muons, is used as direct input to the jet clustering algorithm. As a consequence it is possible to reconstruct jets solely composed by a \jpsi. Solid lines are the predictions from \oldpy using the Monash tune, the red line employs the MPI color reconnection approach while the blue one is simulated with the new more QCD based scheme together with the new beam remnant model \cite{MonashTune}. The simulation is generated with all the color singlet and octet states available in the event generator together with the default LDME. As can be seen, \oldpy results are independent of the CR model and the event generator clearly overshoots the data when \z $\approx$ 1. This is an indication that NRQCD predicts an important fraction of prompt \jpsi to be produced isolated, opposite to the data. A similar result, reported by CMS at \s = 5.02 TeV, is shown in figure \ref{CMSOriginal} \cite{CMSJpsiJet}. Albeit the data analysis procedure followed by CMS is very similar to LHCb, the kinematic range of the measurement in the former experiment is different:  \pt(\jpsi) $>$ 6.5 \gevc, 30 $<$ \pt(jet) $<$ 40 \gevc, $| \eta\mathrm{(jet)} | <$ 2 and jet cone radius of $R=0.3$. Besides this, the \oldpy simulations that are compared to data use the CP5 underlying event tune \cite{CP5}. As a result of the requirement on the \jpsi \pt, due to the acceptance of the detector, the measurements in CMS are limited down to \z $>$ 0.22. Once again the MPI and QCD based color reconnection approaches follow the same behavior, but now there is a striking difference between data and simulation, the later indicating that \jpsi are produced with very few jet activity. According to the simulations, at both experiments the dominant \qqbar pair production mode is given by gluon fusion, followed by flavour excitation and a negligible contribution from light quark and anti-quark annihilation. 

The reason why \oldpy can not reproduce the data can be explained by two factors. The first one is that color singlet states do not radiate during the parton shower process, so the \jpsi is produced isolated. The second factor is relative to the octet states that are allowed to radiate with a splitting probability $Q \rightarrow Q\,g$ multiplied by two, that is equivalent to treat the two heavy quarks as radiators. However there is the quarkonia production by gluon splitting that serves as counterbalance and the net effect leads to \jpsi production with a small degree of jet activity. 

\section{Quarkonia parton shower}\label{sec3}

The inability of \oldpy to correctly describe the experimental results of \jpsi production in jets at the LHC is an indication that quarkonia can be produced through different mechanisms to those described in the previous section. In fact, quarkonia production at different stages of the collision can be modeled by fragmentation functions or parton showers. For this reason there is a new implementation for quarkonia production that introduces quarkonia splittings during the parton shower \cite{OniaShower}. For \jpsi and \ups the available splitting kernels are $Q \rightarrow Q\bar{Q}[^3S_1^{(1)}]Q$, $g \rightarrow Q\bar{Q}[^3S_1^{(1)}]gg$ and $g \rightarrow Q\bar{Q}[^3S_1^{(8)}]$. In this new quarkonia parton shower the only color octet state implemented is treated as a massive gluon. Compared to the LDME in \py it is possible to notice some missing splitting kernels: the gluon and heavy quark initiated to color octet states. On one hand these gluon initiated splittings enter at an order of $\alpha_{\mathrm{s}}^2$, while the $g \rightarrow Q\bar{Q}[^3S_1^{(8)}]$ corresponds to $\alpha_{\mathrm{s}}$. On the other hand, the heavy quark initiated splittings are suppressed relative to the corresponding color singlet splittings.

\begin{figure}
  \begin{subfigure}{0.49\textwidth}
    \includegraphics[width=\linewidth]{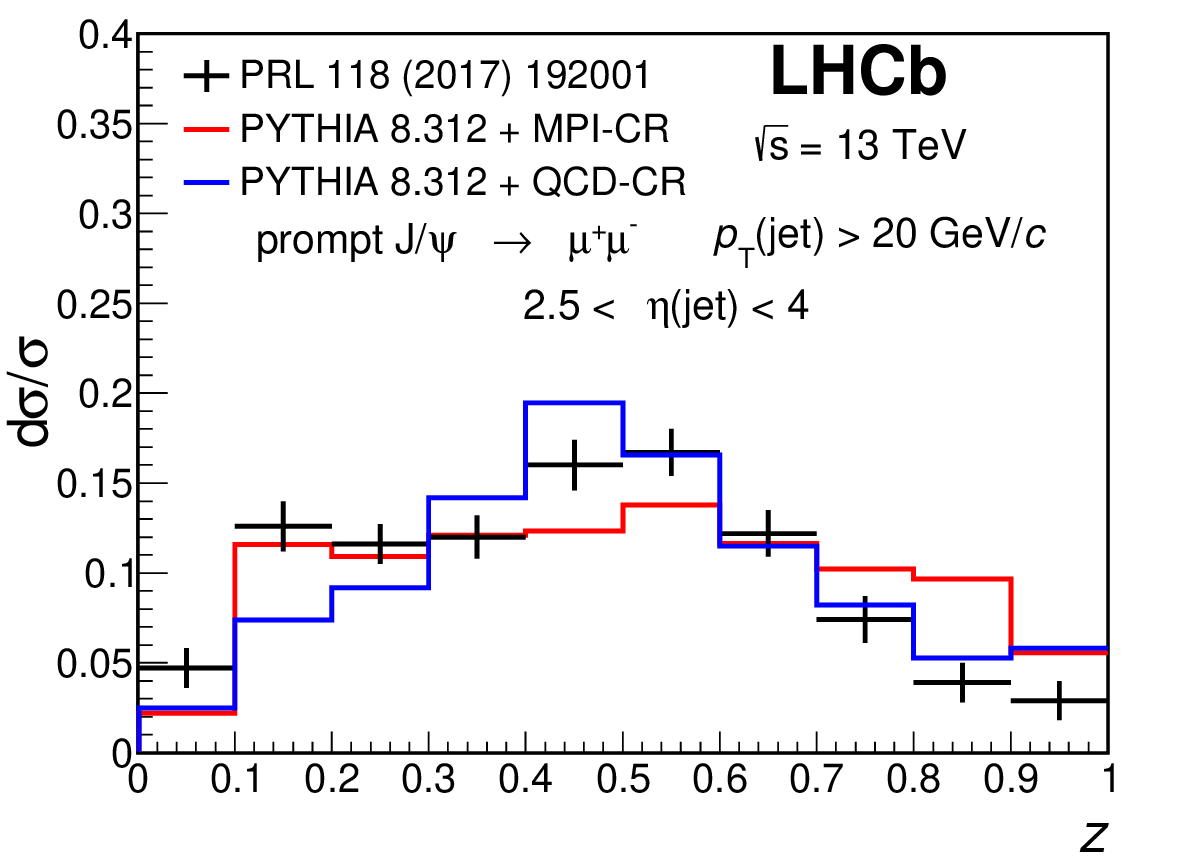}
    \caption{} \label{LHCbNew}
  \end{subfigure}
  \hspace*{-0.4cm}
  \begin{subfigure}{0.49\textwidth}
    \includegraphics[width=\linewidth]{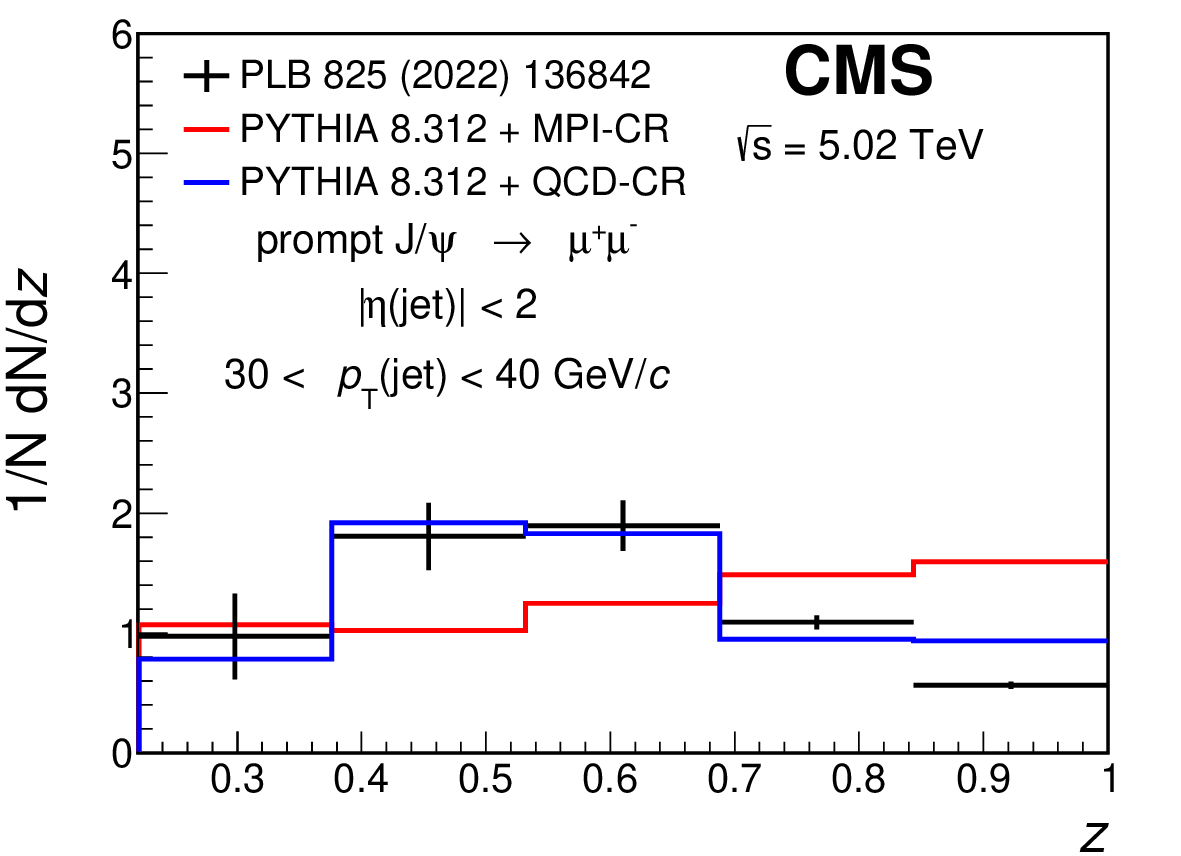}
    \caption{} \label{CMSnew}
  \end{subfigure}
\caption{Normalized \z distributions for prompt \jpsi in pp collisions at \s = 13 and  5.02 TeV as reported by LHCb (a) and CMS (b), respectively. For LHCb the error bars are systematic uncertainties only, while for CMS the statistical and systematic uncertainties have been added in quadrature. Solid lines are the predictions from PYTHIA 8.312, red line includes the new quarkonia parton shower together with the MPI color reconnection approach while the blue one replaces the MPI with the QCD CR based scheme} 
\label{JpsiNew}
\end{figure}

Figure \ref{JpsiNew} shows, once again the normalized \z distributions for prompt \jpsi in pp collisions at \s = 13 and  5.02 TeV as reported by LHCb (a) and CMS (b). The solid lines are now computed with \newpy that includes the new quarkonia parton shower. The only difference is that the red line employs the MPI color reconnection approach while the blue one uses the QCD based scheme. Comparing figure \ref{LHCbNew} to figure \ref{LHCbOriginal}, it is clear that there is a change on the distributions. As can be seen, the new quarkonia shower with MPI-CR tends to flatten out the distribution, making it basically constant for $0.1 < z < 0.7$. The simulation is now in better agreement with the data, showing that the inclusion of the quarkonia splitting kernels during the parton shower provide the correct amount of jet activity to \jpsi production. Replacing the MPI-CR model with the QCD based one provides a stepped distribution that also reproduces the experimental results. Now, concerning CMS in figure \ref{CMSnew}, and compared to the corresponding plot in figure \ref{OriginalPlots}, indicates that including the quarkonia shower generates a dramatic change in the simulations. As a matter of fact the largely favored isolated \jpsi production has now disappeared, albeit the red curve shows an increasing trend when \z approaches its maximum value. This behavior does not appear when the QCD model for color reconnection is used as it can correctly predict the data. As can be seen for CMS, there is now a sizable difference between the two simulations, so the data clearly favours the QCD-CR model applied together with the quarkonia parton shower. 

\begin{figure}
  \begin{subfigure}{0.49\textwidth}
    \includegraphics[width=\linewidth]{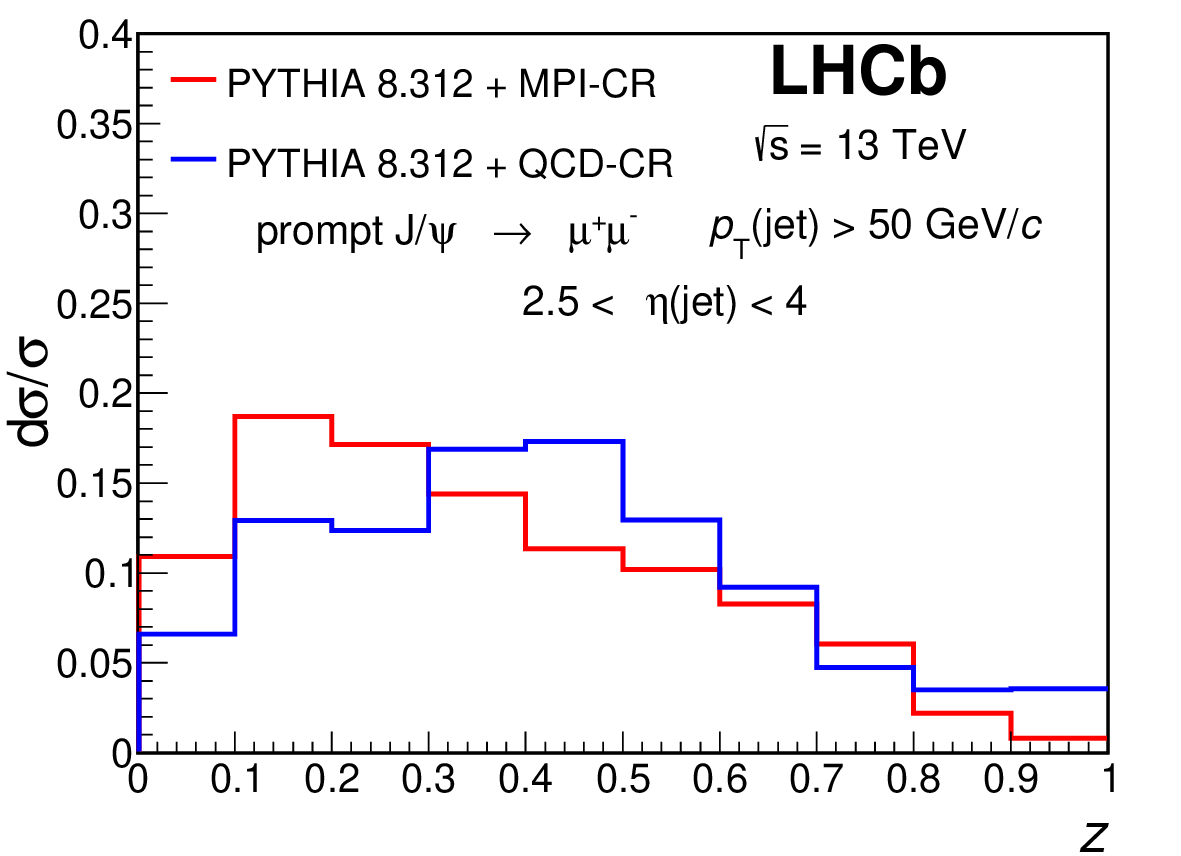}
    \caption{} \label{LHCbFuture}
  \end{subfigure}
  \hspace*{-0.4cm}
  \begin{subfigure}{0.49\textwidth}
    \includegraphics[width=\linewidth]{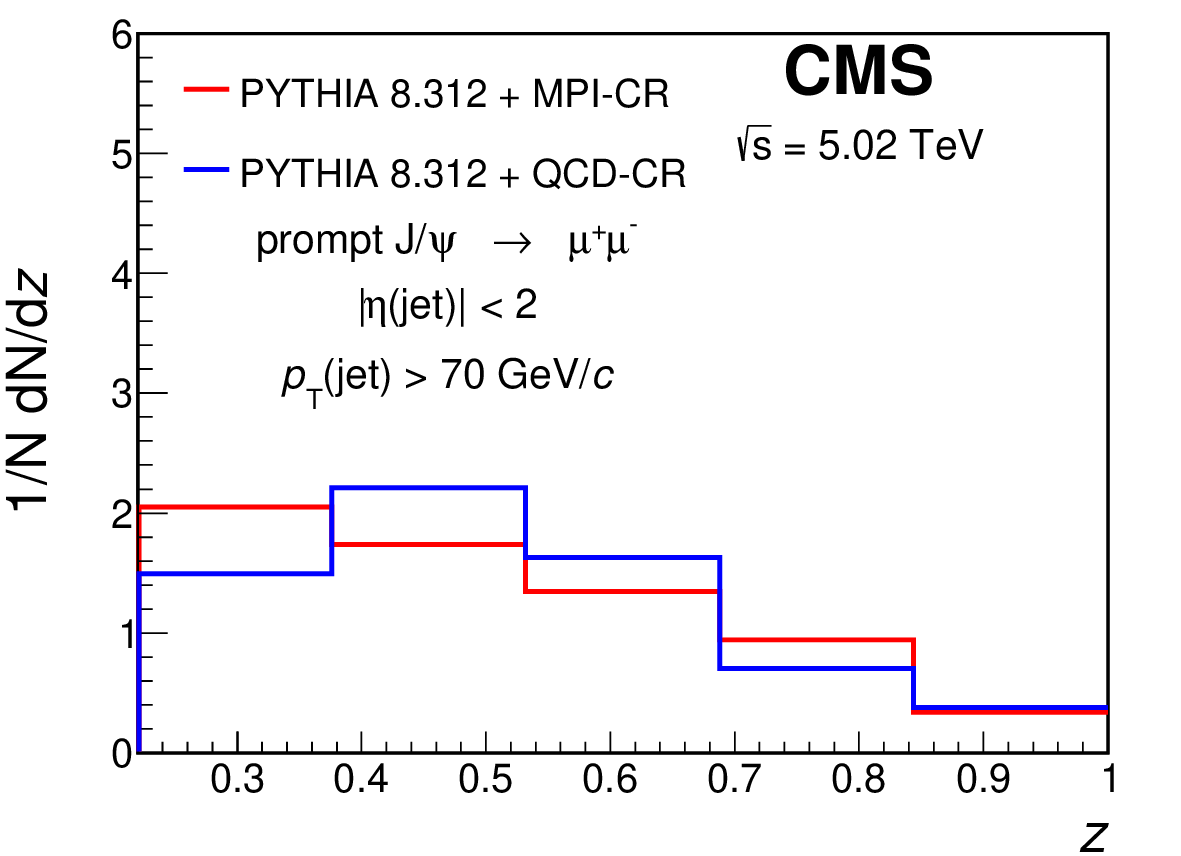}
    \caption{} \label{CMSfuture}
  \end{subfigure}
\caption{Normalized \z distributions predicted for prompt \jpsi in pp collisions at \s = 13 and 5.02 TeV for LHCb (a) and CMS (b), respectively. The jet \pt cuts have now been increased to 50 \gevc for LHCb and 70 \gevc for CMS} 
\label{JpsiFuture}
\end{figure}

The previous effect can be used to test the two color reconnection models. Figure \ref{JpsiFuture} shows the predictions from \newpy for LHCb (a) and CMS (b) if the jet \pt cut is increased to 50 and 70 \gevc, respectively. For LHCb there is now a different behavior depending if MPI or QCD based color reconnection models are used. The former peaks at lower \z values and decreases smoothly towards \z $\approx$ 1. The QCD-CR model still has a similar shape as when the jet \pt cut was set to 20 \gevc. The difference between both distributions for \z $\lesssim$ 0.6 could be used to discriminate between the two CR models. For CMS, discrimination among CR approaches will only be possible for \z $\lesssim$ 0.5 values.

\begin{figure}
  \begin{subfigure}{0.49\textwidth}
    \includegraphics[width=\linewidth]{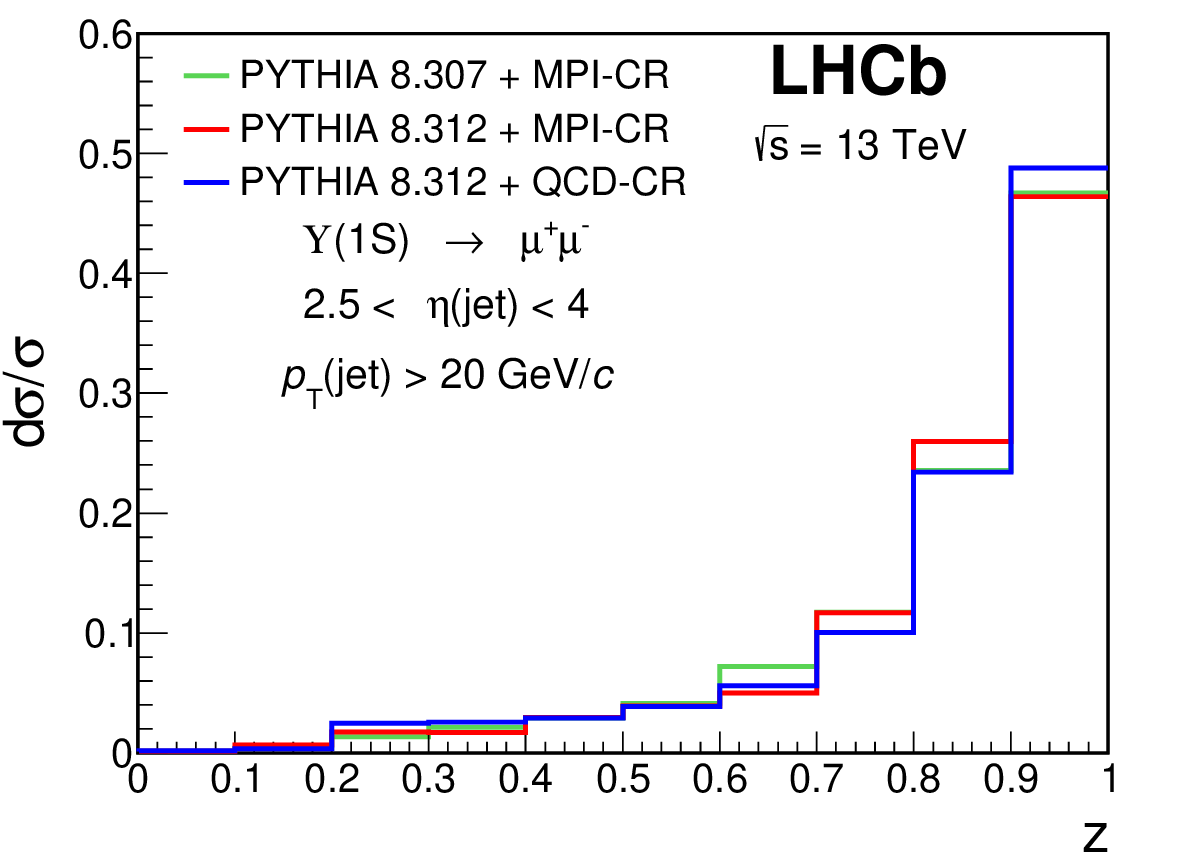}
    \caption{} \label{LHCbUps20}
  \end{subfigure}
  \hspace*{-0.4cm}
  \begin{subfigure}{0.49\textwidth}
    \includegraphics[width=\linewidth]{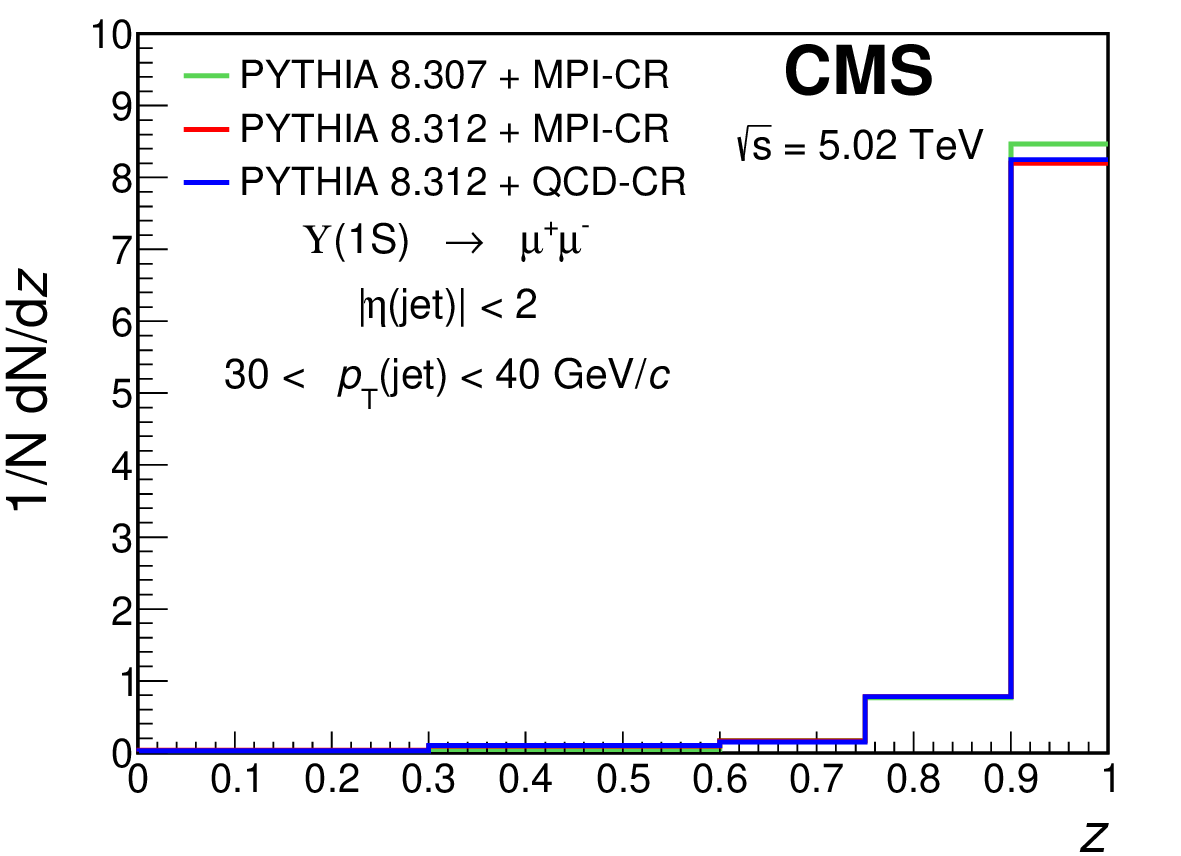}
    \caption{} \label{CMSups3040}
  \end{subfigure}
\caption{Normalized \z distributions predicted for \ups in pp collisions at \s = 13 and 5.02 TeV for LHCb (a) and CMS (b), respectively. Same jet \pt cuts are applied as in the case of the \jpsi. Green distribution is generated with \oldpy using the MPI-CR method. Red and blue distributions are generated with \newpy plus MPI and QCD color reconnection, respectively} 
\label{Upsilon1}
\end{figure}

\begin{figure}
  \begin{subfigure}{0.49\textwidth}
    \includegraphics[width=\linewidth]{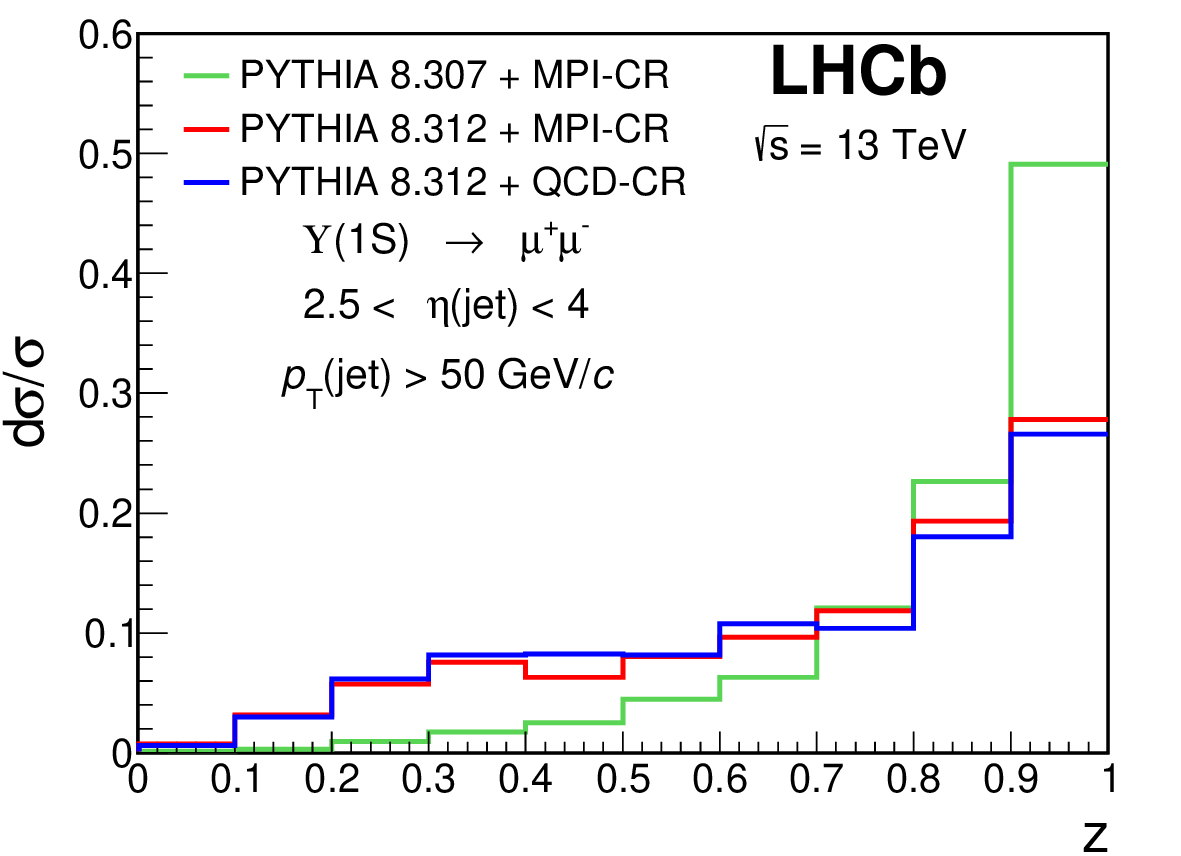}
    \caption{} \label{LHCbUps50}
  \end{subfigure}
  \hspace*{-0.4cm}
  \begin{subfigure}{0.49\textwidth}
    \includegraphics[width=\linewidth]{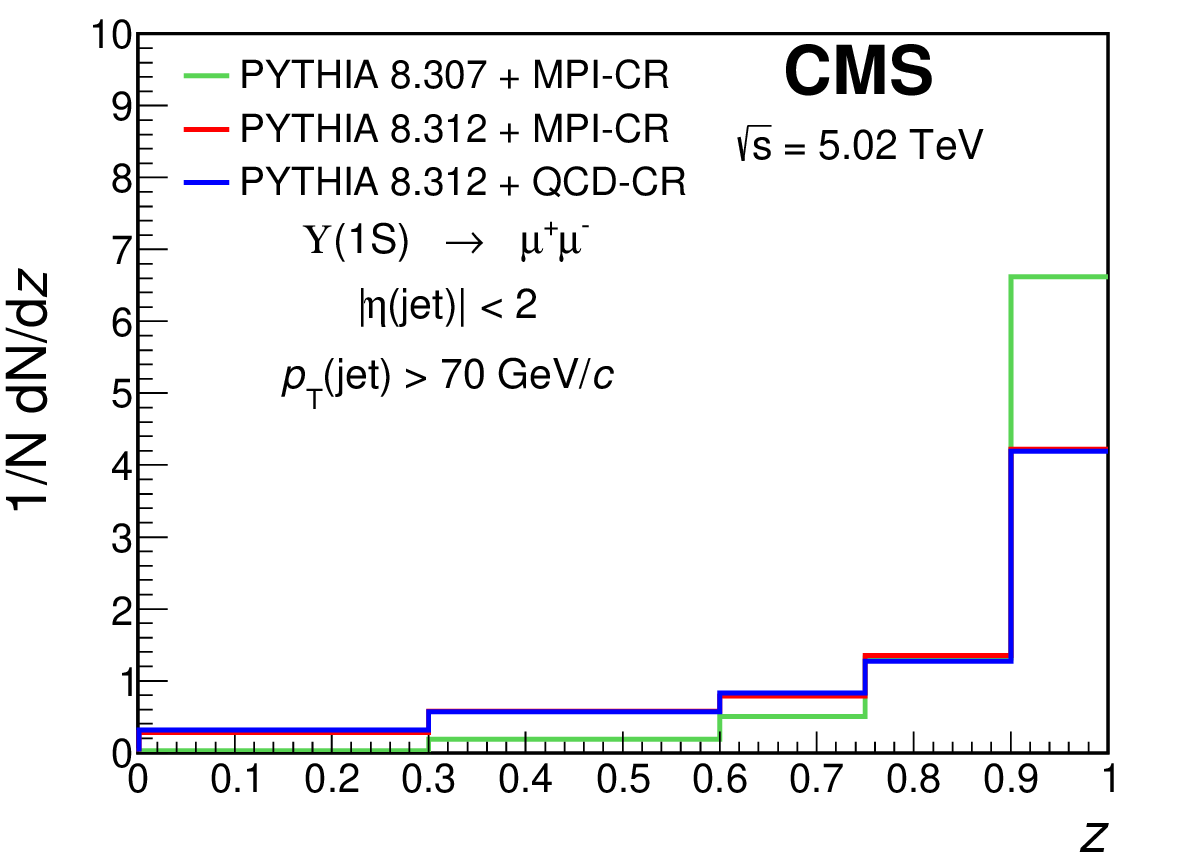}
    \caption{} \label{CMSups70}
  \end{subfigure}
\caption{Normalized \z distributions predicted for \ups in pp collisions at \s = 13 and 5.02 TeV for LHCb (a) and CMS (b), respectively. The jet \pt cuts have now been increased to 50 \gevc for LHCb and 70 \gevc for CMS} 
\label{Upsilon2}
\end{figure}

Once the new quarkonia parton shower has been tested on actual data, it can be used to make predictions on future measurements. Concerning a $^{3}\mathrm{S}_1$ bottomonium state, the most natural choice is \ups. The first exercise is performed in exactly the same kinematic range as it was previously done for the \jpsi. The only difference is that now the track selection is slightly modified according to existing analysis on \ups measurements by LHCb and CMS \cite{LHCbUpsilon,CMSUpsilon3S}. In particular for CMS there is no further restriction on the transverse momentum of the \ups, so it can be measured down to \pt $>$ 0 and as a consequence there is no longer constraint on the lower limit for \z. For LHCb the same binning is used as in the \jpsi case, while for CMS it has been changed. Figure \ref{Upsilon1} shows the predictions from the two different versions of the Monte Carlo used. For \oldpy the same results are obtained wether MPI or QCD based color reconnection is used. For this reason, and for better visualization, only one CR mode is shown (green color). The first evident property is that the event generator predicts for CMS  virtually all the \ups to be produced isolated, while for LHCb the bottomonium state will have some jet activity. The second property observed is that, opposite to the \jpsi case, there is almost no difference between the results depicted by \oldpy and \newpy as all the curves are overlapped. As a consequence, it would not be possible to discriminate between predictions from the different PYTHIA 8 versions. However, this is not a flaw from the quarkonia parton shower and the key point is the mass difference between the \jpsi and the \ups. Indeed, being the later around three time more massive requires a parton with higher energy to create a \ups during the shower. Such effect can be tested by significantly increasing the \pt cut on the jet. This exercise is performed in figure \ref{Upsilon2}, where the new lower \pt thresholds for the jets are now set as 50 and 70 \gevc for LHCb and CMS, respectively. The effect is more notorious when \z $\approx$ 1. As can be seen, at this kinematic regime it will be possible to disentangle the predictions with and without the quarkonia shower.

\section{Conclusion}\label{sec13}

A study on prompt \jpsi and \ups production in jets at LHCb ($\sqrt{s}$ = 13 TeV) and CMS ($\sqrt{s}$ = 5.02 TeV) has been performed using \oldpy and \newpy. Results from the former version can not describe the experimental data of both experiments, predicting that a significant amount of the prompt \jpsi production is isolated. This remains true wether the MPI or QCD based color reconnection model is used. If the new quarkonia parton shower present in \newpy is used, then the agreement with experimental data shows a dramatic improvement, in particular when the QCD-CR is used. This opens the possibility to use the fragmentation of jets containing a prompt \jpsi to discriminate among MPI and QCD color reconnection models, even if the jet \pt selection cut is increased. Finally a prediction for \ups with the same jet \pt requirement as in the prompt \jpsi case indicates that it will be impossible to distinguish between PYTHIA 8 simulations with and without the quarkonia parton shower, independently if the MPI or QCD color reconnection is used. Such distinction can only be made by increasing the jet \pt cut to, at least, 50 \gevc for LHCb and 70 \gevc for CMS.

\bmhead{Acknowledgements}

This work has employed an important amount of computing resources from the ACARUS at the Universidad de Sonora, without this facility it would have been impossible to develop this study.

\bmhead{Data Availability Statement}

This manuscript has no associated data as the results are based on simulations. The event generator employed is publicly available.

\bibliography{sn-bibliography}


\begin{thebibliography}{30}
\ifx \bisbn   \undefined \def \bisbn  #1{ISBN #1}\fi
\ifx \binits  \undefined \def \binits#1{#1}\fi
\ifx \bauthor  \undefined \def \bauthor#1{#1}\fi
\ifx \batitle  \undefined \def \batitle#1{#1}\fi
\ifx \bjtitle  \undefined \def \bjtitle#1{#1}\fi
\ifx \bvolume  \undefined \def \bvolume#1{\textbf{#1}}\fi
\ifx \byear  \undefined \def \byear#1{#1}\fi
\ifx \bissue  \undefined \def \bissue#1{#1}\fi
\ifx \bfpage  \undefined \def \bfpage#1{#1}\fi
\ifx \blpage  \undefined \def \blpage #1{#1}\fi
\ifx \burl  \undefined \def \burl#1{\textsf{#1}}\fi
\ifx \doiurl  \undefined \def \doiurl#1{\url{https://doi.org/#1}}\fi
\ifx \betal  \undefined \def \betal{\textit{et al.}}\fi
\ifx \binstitute  \undefined \def \binstitute#1{#1}\fi
\ifx \binstitutionaled  \undefined \def \binstitutionaled#1{#1}\fi
\ifx \bctitle  \undefined \def \bctitle#1{#1}\fi
\ifx \beditor  \undefined \def \beditor#1{#1}\fi
\ifx \bpublisher  \undefined \def \bpublisher#1{#1}\fi
\ifx \bbtitle  \undefined \def \bbtitle#1{#1}\fi
\ifx \bedition  \undefined \def \bedition#1{#1}\fi
\ifx \bseriesno  \undefined \def \bseriesno#1{#1}\fi
\ifx \blocation  \undefined \def \blocation#1{#1}\fi
\ifx \bsertitle  \undefined \def \bsertitle#1{#1}\fi
\ifx \bsnm \undefined \def \bsnm#1{#1}\fi
\ifx \bsuffix \undefined \def \bsuffix#1{#1}\fi
\ifx \bparticle \undefined \def \bparticle#1{#1}\fi
\ifx \barticle \undefined \def \barticle#1{#1}\fi
\bibcommenthead
\ifx \bconfdate \undefined \def \bconfdate #1{#1}\fi
\ifx \botherref \undefined \def \botherref #1{#1}\fi
\ifx \url \undefined \def \url#1{\textsf{#1}}\fi
\ifx \bchapter \undefined \def \bchapter#1{#1}\fi
\ifx \bbook \undefined \def \bbook#1{#1}\fi
\ifx \bcomment \undefined \def \bcomment#1{#1}\fi
\ifx \oauthor \undefined \def \oauthor#1{#1}\fi
\ifx \citeauthoryear \undefined \def \citeauthoryear#1{#1}\fi
\ifx \endbibitem  \undefined \def \endbibitem {}\fi
\ifx \bconflocation  \undefined \def \bconflocation#1{#1}\fi
\ifx \arxivurl  \undefined \def \arxivurl#1{\textsf{#1}}\fi
\csname PreBibitemsHook\endcsname

\bibitem[\protect\citeauthoryear{Collins et~al.}{1986}]{FactorizationTheo}
\begin{barticle}
\bauthor{\bsnm{Collins}, \binits{J.C.}},
\bauthor{\bsnm{Soper}, \binits{D.E.}},
\bauthor{\bsnm{Sterman}, \binits{G.F.}}:
\batitle{{Heavy Particle Production in High-Energy Hadron Collisions}}.
\bjtitle{Nucl. Phys. B}
\bvolume{263},
\bfpage{37}
(\byear{1986})
\doiurl{10.1016/0550-3213(86)90026-X}
\end{barticle}
\endbibitem

\bibitem[\protect\citeauthoryear{Einhorn and Ellis}{1975}]{CSM0}
\begin{barticle}
\bauthor{\bsnm{Einhorn}, \binits{M.B.}},
\bauthor{\bsnm{Ellis}, \binits{S.D.}}:
\batitle{Hadronic production of the new resonances: Probing gluon
  distributions}.
\bjtitle{Phys. Rev. D}
\bvolume{12},
\bfpage{2007}--\blpage{2014}
(\byear{1975})
\doiurl{10.1103/PhysRevD.12.2007}
\end{barticle}
\endbibitem

\bibitem[\protect\citeauthoryear{Carlson and Suaya}{1976}]{CSM1}
\begin{barticle}
\bauthor{\bsnm{Carlson}, \binits{C.E.}},
\bauthor{\bsnm{Suaya}, \binits{R.}}:
\batitle{Hadronic production of the $\frac{\ensuremath{\psi}}{J}$ meson}.
\bjtitle{Phys. Rev. D}
\bvolume{14},
\bfpage{3115}--\blpage{3121}
(\byear{1976})
\doiurl{10.1103/PhysRevD.14.3115}
\end{barticle}
\endbibitem

\bibitem[\protect\citeauthoryear{Ellis et~al.}{1976}]{CSM2}
\begin{barticle}
\bauthor{\bsnm{Ellis}, \binits{S.D.}},
\bauthor{\bsnm{Einhorn}, \binits{M.B.}},
\bauthor{\bsnm{Quigg}, \binits{C.}}:
\batitle{Comment on hadronic production of psions}.
\bjtitle{Phys. Rev. Lett.}
\bvolume{36},
\bfpage{1263}--\blpage{1266}
(\byear{1976})
\doiurl{10.1103/PhysRevLett.36.1263}
\end{barticle}
\endbibitem

\bibitem[\protect\citeauthoryear{Fritzsch}{1977}]{CEM1}
\begin{barticle}
\bauthor{\bsnm{Fritzsch}, \binits{H.}}:
\batitle{Producing heavy quark flavors in hadronic collisions—' a test of
  quantum chromodynamics}.
\bjtitle{Physics Letters B}
\bvolume{67}(\bissue{2}),
\bfpage{217}--\blpage{221}
(\byear{1977})
\doiurl{10.1016/0370-2693(77)90108-3}
\end{barticle}
\endbibitem

\bibitem[\protect\citeauthoryear{Halzen}{1977}]{CEM2}
\begin{barticle}
\bauthor{\bsnm{Halzen}, \binits{F.}}:
\batitle{Cvc for gluons and hadroproduction of quark flavours}.
\bjtitle{Physics Letters B}
\bvolume{69}(\bissue{1}),
\bfpage{105}--\blpage{108}
(\byear{1977})
\doiurl{10.1016/0370-2693(77)90144-7}
\end{barticle}
\endbibitem

\bibitem[\protect\citeauthoryear{Amundson et~al.}{1996}]{CEM3}
\begin{barticle}
\bauthor{\bsnm{Amundson}, \binits{J.F.}},
\bauthor{\bsnm{Eboli}, \binits{O.J.P.}},
\bauthor{\bsnm{Gregores}, \binits{E.M.}},
\bauthor{\bsnm{Halzen}, \binits{F.}}:
\batitle{{Colorless states in perturbative QCD: Charmonium and rapidity gaps}}.
\bjtitle{Phys. Lett. B}
\bvolume{372},
\bfpage{127}--\blpage{132}
(\byear{1996})
\doiurl{10.1016/0370-2693(96)00035-4}
{\href{https://arxiv.org/abs/hep-ph/9512248}{{arXiv:hep-ph/9512248}}}
\end{barticle}
\endbibitem

\bibitem[\protect\citeauthoryear{Caswell and Lepage}{1986}]{NRQCD1}
\begin{barticle}
\bauthor{\bsnm{Caswell}, \binits{W.E.}},
\bauthor{\bsnm{Lepage}, \binits{G.P.}}:
\batitle{{Effective Lagrangians for Bound State Problems in QED, QCD, and Other
  Field Theories}}.
\bjtitle{Phys. Lett. B}
\bvolume{167},
\bfpage{437}--\blpage{442}
(\byear{1986})
\doiurl{10.1016/0370-2693(86)91297-9}
\end{barticle}
\endbibitem

\bibitem[\protect\citeauthoryear{Bodwin et~al.}{1995}]{NRQCD2}
\begin{barticle}
\bauthor{\bsnm{Bodwin}, \binits{G.T.}},
\bauthor{\bsnm{Braaten}, \binits{E.}},
\bauthor{\bsnm{Lepage}, \binits{G.P.}}:
\batitle{{Rigorous QCD analysis of inclusive annihilation and production of
  heavy quarkonium}}.
\bjtitle{Phys. Rev. D}
\bvolume{51},
\bfpage{1125}--\blpage{1171}
(\byear{1995})
\doiurl{10.1103/PhysRevD.55.5853}
{\href{https://arxiv.org/abs/hep-ph/9407339}{{arXiv:hep-ph/9407339}}}.
\bcomment{[Erratum: Phys.Rev.D 55, 5853 (1997)]}
\end{barticle}
\endbibitem

\bibitem[\protect\citeauthoryear{Sirunyan et~al.}{2018}]{CMSquarkonia}
\begin{barticle}
\bauthor{\bsnm{Sirunyan}, \binits{A.M.}}, \betal:
\batitle{{Measurement of quarkonium production cross sections in pp collisions
  at $\sqrt{s}=$ 13 TeV}}.
\bjtitle{Phys. Lett. B}
\bvolume{780},
\bfpage{251}--\blpage{272}
(\byear{2018})
\doiurl{10.1016/j.physletb.2018.02.033}
{\href{https://arxiv.org/abs/1710.11002}{{arXiv:1710.11002}}}
{[hep-ex]}
\end{barticle}
\endbibitem

\bibitem[\protect\citeauthoryear{Aaij et~al.}{2021}]{LHCbJpsi}
\begin{barticle}
\bauthor{\bsnm{Aaij}, \binits{R.}}, \betal:
\batitle{{Measurement of $J/\psi$ production cross-sections in $pp$ collisions
  at $\sqrt{s}=5$ TeV}}.
\bjtitle{JHEP}
\bvolume{11},
\bfpage{181}
(\byear{2021})
\doiurl{10.1007/JHEP11(2021)181}
{\href{https://arxiv.org/abs/2109.00220}{{arXiv:2109.00220}}}
{[hep-ex]}
\end{barticle}
\endbibitem

\bibitem[\protect\citeauthoryear{Aaij et~al.}{2018}]{LHCbUpsilon}
\begin{barticle}
\bauthor{\bsnm{Aaij}, \binits{R.}}, \betal:
\batitle{{Measurement of $\Upsilon$ production in $pp$ collisions at
  $\sqrt{s}$= 13 TeV}}.
\bjtitle{JHEP}
\bvolume{07},
\bfpage{134}
(\byear{2018})
\doiurl{10.1007/JHEP07(2018)134}
{\href{https://arxiv.org/abs/1804.09214}{{arXiv:1804.09214}}}
{[hep-ex]}.
\bcomment{[Erratum: JHEP 05, 076 (2019)]}
\end{barticle}
\endbibitem

\bibitem[\protect\citeauthoryear{Acharya et~al.}{2023}]{ALICEquarkonium}
\begin{barticle}
\bauthor{\bsnm{Acharya}, \binits{S.}}, \betal:
\batitle{{Inclusive quarkonium production in pp collisions at $\sqrt{s}$ =
  5.02~TeV}}.
\bjtitle{Eur. Phys. J. C}
\bvolume{83}(\bissue{1}),
\bfpage{61}
(\byear{2023})
\doiurl{10.1140/epjc/s10052-022-10896-8}
{\href{https://arxiv.org/abs/2109.15240}{{arXiv:2109.15240}}}
{[nucl-ex]}
\end{barticle}
\endbibitem

\bibitem[\protect\citeauthoryear{Aad et~al.}{2024}]{ATLASjpsi}
\begin{barticle}
\bauthor{\bsnm{Aad}, \binits{G.}}, \betal:
\batitle{{Measurement of the production cross-section of $J/\psi $ and $\psi
  (2{\textrm{S}})$ mesons in pp collisions at $\sqrt{s} = 13$~TeV with the
  ATLAS detector}}.
\bjtitle{Eur. Phys. J. C}
\bvolume{84}(\bissue{2}),
\bfpage{169}
(\byear{2024})
\doiurl{10.1140/epjc/s10052-024-12439-9}
{\href{https://arxiv.org/abs/2309.17177}{{arXiv:2309.17177}}}
{[hep-ex]}
\end{barticle}
\endbibitem

\bibitem[\protect\citeauthoryear{Hayrapetyan et~al.}{2024}]{CMSpolJpsi}
\begin{barticle}
\bauthor{\bsnm{Hayrapetyan}, \binits{A.}}, \betal:
\batitle{{Measurement of the polarizations of prompt and non-prompt Image 1 and
  \ensuremath{\psi}(2S) mesons produced in pp collisions at s=13TeV}}.
\bjtitle{Phys. Lett. B}
\bvolume{858},
\bfpage{139044}
(\byear{2024})
\doiurl{10.1016/j.physletb.2024.139044}
{\href{https://arxiv.org/abs/2406.14409}{{arXiv:2406.14409}}}
{[hep-ex]}
\end{barticle}
\endbibitem

\bibitem[\protect\citeauthoryear{Aaij et~al.}{2017}]{LHCbPolUps}
\begin{barticle}
\bauthor{\bsnm{Aaij}, \binits{R.}}, \betal:
\batitle{{Measurement of the $\Upsilon$ polarizations in $pp$ collisions at
  $\sqrt{s}=7$ and 8 TeV}}.
\bjtitle{JHEP}
\bvolume{12},
\bfpage{110}
(\byear{2017})
\doiurl{10.1007/JHEP12(2017)110}
{\href{https://arxiv.org/abs/1709.01301}{{arXiv:1709.01301}}}
{[hep-ex]}
\end{barticle}
\endbibitem

\bibitem[\protect\citeauthoryear{Acharya et~al.}{2022}]{ALICEjpsiMult}
\begin{barticle}
\bauthor{\bsnm{Acharya}, \binits{S.}}, \betal:
\batitle{{Forward rapidity J/\ensuremath{\psi} production as a function of
  charged-particle multiplicity in pp collisions at $ \sqrt{s} $ = 5.02 and 13
  TeV}}.
\bjtitle{JHEP}
\bvolume{06},
\bfpage{015}
(\byear{2022})
\doiurl{10.1007/JHEP06(2022)015}
{\href{https://arxiv.org/abs/2112.09433}{{arXiv:2112.09433}}}
{[nucl-ex]}
\end{barticle}
\endbibitem

\bibitem[\protect\citeauthoryear{Sirunyan et~al.}{2020}]{CMSupsMult}
\begin{barticle}
\bauthor{\bsnm{Sirunyan}, \binits{A.M.}}, \betal:
\batitle{{Investigation into the event-activity dependence of $\Upsilon$(nS)
  relative production in proton-proton collisions at $ \sqrt{s} $ = 7 TeV}}.
\bjtitle{JHEP}
\bvolume{11},
\bfpage{001}
(\byear{2020})
\doiurl{10.1007/JHEP11(2020)001}
{\href{https://arxiv.org/abs/2007.04277}{{arXiv:2007.04277}}}
{[hep-ex]}
\end{barticle}
\endbibitem

\bibitem[\protect\citeauthoryear{}{2022}]{ATLASupsMult}
\begin{botherref}
{Correlation of $\Upsilon$ meson production with the underlying event in $pp$
  collisions measured by the ATLAS experiment}.
Technical report,
CERN,
Geneva
(2022).
All figures including auxiliary figures are available at
  https://atlas.web.cern.ch/Atlas/GROUPS/PHYSICS/CONFNOTES/ATLAS-CONF-2022-023.
\url{https://cds.cern.ch/record/2806464}
\end{botherref}
\endbibitem

\bibitem[\protect\citeauthoryear{Aaij et~al.}{2017}]{LHCbJpsiJet}
\begin{barticle}
\bauthor{\bsnm{Aaij}, \binits{R.}}, \betal:
\batitle{{Study of J/\ensuremath{\psi} Production in Jets}}.
\bjtitle{Phys. Rev. Lett.}
\bvolume{118}(\bissue{19}),
\bfpage{192001}
(\byear{2017})
\doiurl{10.1103/PhysRevLett.118.192001}
{\href{https://arxiv.org/abs/1701.05116}{{arXiv:1701.05116}}}
{[hep-ex]}
\end{barticle}
\endbibitem

\bibitem[\protect\citeauthoryear{Tumasyan et~al.}{2022}]{CMSJpsiJet}
\begin{barticle}
\bauthor{\bsnm{Tumasyan}, \binits{A.}}, \betal:
\batitle{{Fragmentation of jets containing a prompt J$/\psi$ meson in PbPb and
  pp collisions at $\sqrt{s_\mathrm{NN}} =$ 5.02 TeV}}.
\bjtitle{Phys. Lett. B}
\bvolume{825},
\bfpage{136842}
(\byear{2022})
\doiurl{10.1016/j.physletb.2021.136842}
{\href{https://arxiv.org/abs/2106.13235}{{arXiv:2106.13235}}}
{[hep-ex]}
\end{barticle}
\endbibitem

\bibitem[\protect\citeauthoryear{Bierlich et~al.}{2022}]{Pythia8.3}
\begin{barticle}
\bauthor{\bsnm{Bierlich}, \binits{C.}}, \betal:
\batitle{{A comprehensive guide to the physics and usage of PYTHIA 8.3}}.
\bjtitle{SciPost Phys. Codeb.}
\bvolume{2022},
\bfpage{8}
(\byear{2022})
\doiurl{10.21468/SciPostPhysCodeb.8}
{\href{https://arxiv.org/abs/2203.11601}{{arXiv:2203.11601}}}
{[hep-ph]}
\end{barticle}
\endbibitem

\bibitem[\protect\citeauthoryear{Christiansen and Skands}{2015}]{QCD-basedCR}
\begin{barticle}
\bauthor{\bsnm{Christiansen}, \binits{J.R.}},
\bauthor{\bsnm{Skands}, \binits{P.Z.}}:
\batitle{{String Formation Beyond Leading Colour}}.
\bjtitle{JHEP}
\bvolume{08},
\bfpage{003}
(\byear{2015})
\doiurl{10.1007/JHEP08(2015)003}
{\href{https://arxiv.org/abs/1505.01681}{{arXiv:1505.01681}}}
{[hep-ph]}
\end{barticle}
\endbibitem

\bibitem[\protect\citeauthoryear{Sjostrand and
  Skands}{2005}]{TimeLikePartonShower}
\begin{barticle}
\bauthor{\bsnm{Sjostrand}, \binits{T.}},
\bauthor{\bsnm{Skands}, \binits{P.Z.}}:
\batitle{{Transverse-momentum-ordered showers and interleaved multiple
  interactions}}.
\bjtitle{Eur. Phys. J. C}
\bvolume{39},
\bfpage{129}--\blpage{154}
(\byear{2005})
\doiurl{10.1140/epjc/s2004-02084-y}
{\href{https://arxiv.org/abs/hep-ph/0408302}{{arXiv:hep-ph/0408302}}}
\end{barticle}
\endbibitem

\bibitem[\protect\citeauthoryear{Cacciari et~al.}{2008}]{AntiKtAlgo}
\begin{barticle}
\bauthor{\bsnm{Cacciari}, \binits{M.}},
\bauthor{\bsnm{Salam}, \binits{G.P.}},
\bauthor{\bsnm{Soyez}, \binits{G.}}:
\batitle{{The anti-$k_t$ jet clustering algorithm}}.
\bjtitle{JHEP}
\bvolume{04},
\bfpage{063}
(\byear{2008})
\doiurl{10.1088/1126-6708/2008/04/063}
{\href{https://arxiv.org/abs/0802.1189}{{arXiv:0802.1189}}}
{[hep-ph]}
\end{barticle}
\endbibitem

\bibitem[\protect\citeauthoryear{Cacciari et~al.}{2012}]{FastJet}
\begin{barticle}
\bauthor{\bsnm{Cacciari}, \binits{M.}},
\bauthor{\bsnm{Salam}, \binits{G.P.}},
\bauthor{\bsnm{Soyez}, \binits{G.}}:
\batitle{{FastJet User Manual}}.
\bjtitle{Eur. Phys. J. C}
\bvolume{72},
\bfpage{1896}
(\byear{2012})
\doiurl{10.1140/epjc/s10052-012-1896-2}
{\href{https://arxiv.org/abs/1111.6097}{{arXiv:1111.6097}}}
{[hep-ph]}
\end{barticle}
\endbibitem

\bibitem[\protect\citeauthoryear{Skands et~al.}{2014}]{MonashTune}
\begin{barticle}
\bauthor{\bsnm{Skands}, \binits{P.}},
\bauthor{\bsnm{Carrazza}, \binits{S.}},
\bauthor{\bsnm{Rojo}, \binits{J.}}:
\batitle{{Tuning PYTHIA 8.1: the Monash 2013 Tune}}.
\bjtitle{Eur. Phys. J. C}
\bvolume{74}(\bissue{8}),
\bfpage{3024}
(\byear{2014})
\doiurl{10.1140/epjc/s10052-014-3024-y}
{\href{https://arxiv.org/abs/1404.5630}{{arXiv:1404.5630}}}
{[hep-ph]}
\end{barticle}
\endbibitem

\bibitem[\protect\citeauthoryear{Sirunyan et~al.}{2020}]{CP5}
\begin{barticle}
\bauthor{\bsnm{Sirunyan}, \binits{A.M.}}, \betal:
\batitle{{Extraction and validation of a new set of CMS PYTHIA8 tunes from
  underlying-event measurements}}.
\bjtitle{Eur. Phys. J. C}
\bvolume{80}(\bissue{1}),
\bfpage{4}
(\byear{2020})
\doiurl{10.1140/epjc/s10052-019-7499-4}
{\href{https://arxiv.org/abs/1903.12179}{{arXiv:1903.12179}}}
{[hep-ex]}
\end{barticle}
\endbibitem

\bibitem[\protect\citeauthoryear{Cooke et~al.}{2024}]{OniaShower}
\begin{barticle}
\bauthor{\bsnm{Cooke}, \binits{N.}},
\bauthor{\bsnm{Ilten}, \binits{P.}},
\bauthor{\bsnm{L\"onnblad}, \binits{L.}},
\bauthor{\bsnm{Mrenna}, \binits{S.}}:
\batitle{{Non-relativistic quantum chromodynamics in parton showers}}.
\bjtitle{Eur. Phys. J. C}
\bvolume{84}(\bissue{4}),
\bfpage{432}
(\byear{2024})
\doiurl{10.1140/epjc/s10052-024-12760-3}
{\href{https://arxiv.org/abs/2312.05203}{{arXiv:2312.05203}}}
{[hep-ph]}
\end{barticle}
\endbibitem

\bibitem[\protect\citeauthoryear{Tumasyan et~al.}{2024}]{CMSUpsilon3S}
\begin{barticle}
\bauthor{\bsnm{Tumasyan}, \binits{A.}}, \betal:
\batitle{Observation of the $\mathrm{\ensuremath{\Upsilon}}(3s)$ meson and
  suppression of $\mathrm{\ensuremath{\Upsilon}}$ states in pb-pb collisions at
  $\sqrt{{s}_{NN}}=5.02\text{ }\text{ }\mathrm{TeV}$}.
\bjtitle{Phys. Rev. Lett.}
\bvolume{133},
\bfpage{022302}
(\byear{2024})
\doiurl{10.1103/PhysRevLett.133.022302}
\end{barticle}
\endbibitem

\end{thebibliography}

\end{document}